\documentclass[aps,prd,12pt,preprint,tightenlines,unsortedaddress,
amsfonts,amssymb,amsmath,byrevtex,showpacs,nofootinbib]{revtex4-1}

\usepackage{latexsym}
\usepackage{graphicx}
\usepackage{geometry}
\usepackage{epstopdf}
\usepackage{color}

\usepackage{xcolor}
\usepackage{pdfsync}
\usepackage{fancyhdr}
\usepackage[applemac]{inputenc}
\usepackage[breaklinks,colorlinks,backref]{hyperref}
\hypersetup{
   colorlinks, 
     linktoc=all, 
      linkcolor=magenta,  
    citecolor=magenta,
    urlcolor=magenta}
   \hypersetup{
    citebordercolor=,
    filebordercolor=green,
    linkbordercolor=blue
  }
  \definecolor{hyptxt}{rgb}{0.7, 0.4, 0.9}
\usepackage{bibtopic}

\newcommand{\ud}{\mathrm{d}}
\newcommand{\be}{\begin{equation}}
\newcommand{\ee}{\end{equation}}

\newcommand{\ket}[1]{|\kern.3ex#1\kern.3ex\rangle}
\newcommand{\bra}[1]{\langle\kern.3ex #1 \kern.3ex|}
\newcommand{\scalar}[2]{\langle\kern.3ex #1 \kern.3ex|\kern.3ex#2\kern.3ex\rangle}
\newcommand{\norm}[1]{\|\kern.3ex#1\kern.3ex \|}

\def\ud{\mathrm{d}}

\def\vap{\varpi}

\def\sfM{\mathsf{M}}
\def\sfMps_{\sfM_{\sigma}^{\vap}}

\def\ud{\mathrm{d}}

\def\arcosh{\textrm{arcosh}}
\def\bsb{\boldsymbol{\beta}}
\makeatletter

\begin{document}

\date{\today}

\title{Spectral properties of the quantum Mixmaster universe}

\author{Herv\'{e} Bergeron}
\email{herve.bergeron@u-psud.fr} \affiliation{Univ Paris-Sud, ISMO, UMR 8214 CNRS,
91405 Orsay, France}

\author{Ewa Czuchry}
\email{ewa.czuchry@ncbj.gov.pl} \affiliation{National Centre for Nuclear Research, Ho{\.z}a 69,
00-681 Warszawa, Poland}

\author{Jean-Pierre Gazeau}
\email{gazeau@apc.univ-paris7.fr}
\affiliation{APC, UMR 7164 CNRS, Univ Paris Diderot, Sorbonne Paris Cit\'e, 75205 Paris, France}
\affiliation{Centro Brasileiro de Pesquisas Fisicas
22290-180 - Rio de Janeiro, RJ, Brazil }

\author{Przemys{\l}aw Ma{\l}kiewicz}
\email{przemyslaw.malkiewicz@ncbj.gov.pl}
\affiliation{National Centre for Nuclear Research,  Ho{\.z}a 69,
00-681
Warszawa, Poland}

\begin{abstract}
{ We study the spectral properties of
the anisotropic part of  Hamiltonian entering the quantum dynamics
of the Mixmaster universe.}  We  derive  the explicit asymptotic expressions for the energy spectrum in the limit of large and small volumes of the universe. Then we study the threshold condition between both regimes. Finally we  prove that the spectrum is purely discrete for any volume of the universe. Our results validate and improve the known approximations to the anisotropy potential. They should be useful for any approach to the quantization of the Mixmaster universe.

\end{abstract}
\maketitle

\tableofcontents

\section{Introduction}

Analytical and numerical results  suggest that the dynamics of the Universe on approach to the big-crunch/big-bang singularity is dominated by the time derivatives of the gravitational field \cite{BKL,DG}. Hence, the dynamics at each spatial point becomes ultralocal, oscillatory, chaotic, and  is driven entirely by the gravitational self-energy. These generic features are exemplified by the Mixmaster universe \cite{cwm}, which is a model of spatially homogeneous and anisotropic spacetime with the incorporated Bianchi type IX symmetry. In the context of quantum gravity,  the Mixmaster universe seems to be an ideal tool for testing whether quantization can resolve the problem of classical singularities.

The canonical formalism of the Mixmaster universe in the Misner variables describes the universe in terms of a particle in a 3-dimensional Minkowski spacetime in a potential {representing the spatial curvature of the universe. }   {The anisotropy part of this potential is a non trivial  function of two variables (see \eqref{b9pot}) for which the Schr\"odinger problem is not integrable.}

{The problem of solving the quantum dynamics of the Mixmaster universe is quite involved. It is true for the traditional approaches based on the Wheeler-DeWitt equation (e.g. see \cite{bae15} and references therein) or the Misner reduced phase space \cite{cwm} as well as for the novel approaches like the one developed by the present authors in \cite{qb9a,qb9b,vib}. The common element of all the approaches is the natural split between the isotropic and anisotropic degrees of freedom and the ensuing decomposition of the Hamiltonian. Although, the anisotropic and isotropic dynamics are coupled and ultimately have to be considered together, the knowledge of properties of the non-trivial anisotropic Hamiltonian is crucial for understanding the full dynamics. In this regard, the Mixmaster universe is analogous to molecular systems that admit a natural split between nuclear and electronic degrees of freedom. This feature is essential in our approach.}

{The knowledge of properties of the anisotropic Hamiltonian is a solid starting point for studying the full model, which includes the coupling between the anisotropic and isotropic variables. The details of such a framework depend on the specific quantization of the isotropic Hamiltonian. The dynamics following from the Wheeler-DeWitt equation is known to be singular, whereas the quantization proposed in \cite{qb9a,qb9b,vib} produces an extra repulsive term that replaces the classical singularity with a bounce. In any case, some quantum trajectories may be sufficiently well determined by means of the adiabatic approximations (the Born-Oppenheimer or the Born-Huang) \cite{qb9a,qb9b}. Determination of more elaborate quantum trajectories requires available nonadiabatic methods, e.g. those used  in the context of chemical reaction dynamics \cite{CTBBO}. The key point is that the knowledge of properties of the anisotropic Hamiltonian enables to reduce the dimensionality of the studied equation. Thus, even though its solution ultimately requires numerical simulations, the control over the space of solutions and the qualitative understanding of dynamics are largely enhanced.}

{The usual approximations for the  anisotropy Schr\"odinger spectral problem are the harmonic or the steep wall approximations (\cite{cwm}, for  recent studies see e.g. \cite{kirillov,ds} and references therein). Their respective validities  have never been rigorously studied.  In particular,  they have never been considered in a unified manner as corresponding  to the two extreme regimes of the volume of the universe. Moreover the limit condition separating the two regimes has never been explicitly  given.  Note that the purely discrete spectrum of the two approximations does not imply a purely discrete one for the exact potential for all volumes.}

In the present paper we fill those crucial gaps in the knowledge of the properties of the Bianchi IX anisotropy potential. {For any quantum system} the knowledge of the full spectrum of the Hamiltonian is crucial. For example, the adiabatic approximation can be considered only for the discrete part of the spectrum {of a relevant subsystem}, and only if this discrete part is not embedded into a continuous one. These features were considered by B. Simon in \cite{barrys}. {Therefore the proof that the Bianchi IX  anisotropy spectrum is indeed purely discrete for any volume of the universe is essential. Furthermore the knowledge of the analytical approximations to the spectrum is decisive: for example, a non-adiabatic framework to the Bianchi IX model is studied in \cite{vib}, but the analytical part of the study is limited (harmonic approximation) by the lack of detailed knowledge of the spectrum.  Since our results concern the analytical properties of the anisotropic Shr\"{o}dinger spectrum which is proper to the Bianchi IX geometry, they should be useful for studies of many quantum models of Mixmaster. Nevertheless, the immediate application of our results is to  validate the assumptions underlying the quantum theory of the Mixmaster universe proposed in \cite{vib,qb9a,qb9b}.}

The outline of the paper is as follows. In Sec. II we recall the essential elements of the canonical formalism for the Mixmaster universe and  the anisotropy potential is analysed. Sec. III deals with the asymptotic analysis of the spectrum of the quantum model in two opposite situations corresponding to  large and small volumes of the Universe. In particular, we highlight a unique unitary transformation that allows to study both limits on the same ground. The limit condition separating both regimes is also given. Moreover we improve the steep wall approximation which is widely used in the literature. In Sec. IV we prove that the spectrum associated with the anisotropy potential is purely discrete irrespectively of the size of the universe. We conclude in Sec. V. \footnote{{Throughout the paper we assume $c=1$.}}

\section{Preliminaries}
The line element of the Bianchi type IX model reads:
\begin{equation}
\ud s^2= -{\cal N}^2\ud\tau^2+\sum_ia_i^2(\omega^i)^2\, ,
\end{equation}
where {$\ud \omega_i=\frac{1}{2} \frak{n}\varepsilon_{i}^{\, jk}\omega_j \wedge \omega_k$. The Hamiltonian constraint of the Mixmaster universe in the Misner variables $(\Omega,p_{\Omega},\bsb, \mathbf{p})\in\mathbb{R}^6$ reads \cite{cwm}:
\begin{equation}\label{con}
\mathrm{C}=\frac{{\cal N}e^{-3\Omega}}{24}\left(\frac{2\kappa}{\mathcal{V}_0}\right)^2\left(-p_{\Omega}^2+\mathbf{ p}^2+36\left(\frac{\mathcal{V}_0}{2\kappa}\right)^3\frak{n}^2e^{4\Omega}[V(\bsb)-1]\right)\,,
\end{equation}
where $\bsb:=(\beta_{+},\beta_-)$, $\mathbf{p}:= (p_+,p_-)$, $\mathcal{V}_0=\frac{16\pi^2}{\frak{n}^3}$ is the fiducial volume, $\kappa=8\pi G$ is the gravitational constant, ${\cal N}$ is the nonvanishing lapse function subject to an arbitrary choice. {The anisotropy potential reads:}
\begin{equation}\label{b9pot}
V(\bsb) = \frac{e^{4\beta_+}}{3} \left[\left(2\cosh(2\sqrt{3}\beta_-)-e^{- 6\beta_+}
\right)^2-4\right] +  1 \,.
\end{equation}
{Henceforth}  $\frak{n}=1$ and $2\kappa=\mathcal{V}_0$.} The gravitational Hamiltonian (\ref{con}) resembles the Hamiltonian of a particle in a 3D Minkowski spacetime in a potential arising from the spatial curvature. The spacetime variables have the following cosmological interpretation:
\begin{equation}\Omega=\frac13\ln a_1a_2a_3,~~\beta_+=\frac16\ln\frac{a_1a_2}{a_3^2},~~\beta_-=\frac{1}{2\sqrt{3}}\ln\frac{a_1}{a_2}~.\end{equation}
Hence,  $\Omega$ describes the isotropic part of geometry, {whereas the anisotropic variables $\beta_{\pm}$ describe distortions to  isotropy. }
The Hamiltonian constraint (\ref{con}) can be decomposed as a sum of isotropic and anisotropic parts which read (up to a non-vanishing factor)
\begin{align}\label{condec}
\mathrm{C}=-\mathrm{C}_{iso}+\mathrm{C}_{ani},~~\mathrm{C}_{iso}=p_{\Omega}^2+36e^{4\Omega},~~\mathrm{C}_{ani}=\mathbf{p}^2+36e^{4\Omega}V(\bsb).
\end{align}
\begin{figure}[t]
\begin{tabular}{cc}
\includegraphics[width=0.4\textwidth]{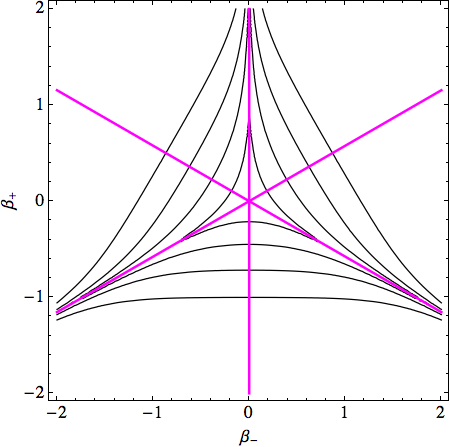} \hspace{1cm}
\includegraphics[width=0.45\textwidth]{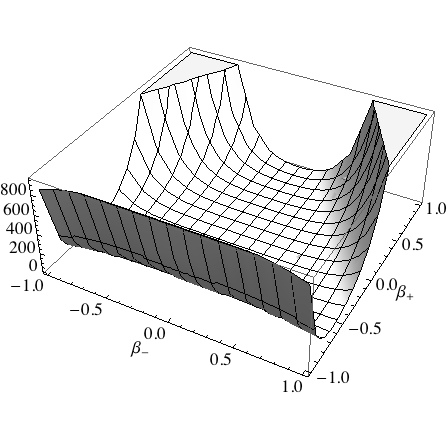}
\end{tabular}
\caption{Plot of the Bianchi type IX anisotropy potential near its minimum with the three ${\sf C}_{3v}$ symmetry axes $\beta_-=0$, $\beta_+=\beta_-/\sqrt{3}$, $\beta_+=-\beta_-/\sqrt{3}$.}
\label{figure1}
\end{figure}
The potential $V$ deserves particular attention due to its three open ${\sf C}_{3v}$ symmetry deep ``canyons'', increasingly narrow until  their respective wall edges close up at the infinity whereas their respective bottoms tend to zero (see Fig. \ref{figure1}). The potential $V$ is asymptotically  {confining} except for these directions in which $V \to 0$:
\begin{align}\nonumber
&\textrm{(i)} ~~~\beta_-=0, ~~\beta_+ \to \infty~,\\
&\textrm{(ii)} ~~ \beta_+= \beta_-/\sqrt{3}, ~~\beta_- \to -\infty~,\\ \nonumber
&\textrm{(iii)}~\beta_+= -\beta_-/\sqrt{3},~~\beta_- \to \infty~.
\end{align}
It is bounded from below and reaches
its absolute minimum value at
$\beta_\pm=0$, where $V=0$. Near its minimum $V$ behaves as the two-dimensional isotropic harmonic potential:
\begin{equation}\label{potharmonic}
V(\bsb) = 8 \bsb^2 +o(\beta_\pm^2)\,.
\end{equation}
Away from its minimum, the so-called steep wall approximation applies as $V$ tends to an equilateral triangle potential with its infinitely steep walls.

We notice in Eq. (\ref{condec}) that during the evolution of the universe towards the singular point, $\Omega\rightarrow -\infty$ and the factor in front of the potential $V$ goes to zero, $36e^{4\Omega}\rightarrow 0$. Therefore, as the universe contracts the potential walls move apart and the particle penetrates larger and larger parts of the anisotropy space $\bsb=(\beta_+,\beta_-)$.

\section{Asymptotic analysis of the spectrum}

{Canonical quantization of the Hamiltonian constraint (\ref{condec}) leads to the well-known Wheeler-DeWitt equation \cite{bae15}. However, as already mentioned, this equation does not remove the classical singularity. A quantization  that removes the singularity (see \cite{qb9a,qb9b} for details) is implemented with the isotropic variables that bring the singular point to finite values, namely:
\begin{equation}
q=\sqrt{a_1a_2a_3},~~p=-\frac{16}{3{\cal N}}\dot{q}\, .
\end{equation}
Then  the full quantum constraint operator reads:
\begin{align}\label{qcon}
\hat{\mathrm{C}}=\frac{\partial^2}{\partial q^2}-\frac{K}{q^2}-36q^{2/3}+\widehat{\mathrm{C}}_{ani}(q),
\end{align}
with $K>0$. The anisotropic part of the quantized (\ref{condec}) reads as the $q$-dependent Schr\"{o}dinger operator acting in the Hilbert space $\mathcal{H}=L^2(\mathbb{R}^2,\ud\beta_+\ud\beta_-)$,
\begin{equation}
\label{quham}
 \widehat{\mathrm{C}}_{ani}(q) := \frac{\hat{\mathbf{p}}^2}{q^2}+
\, q^{2/3}V(\bsb)\,,
\end{equation}
where $\hat{p}_\pm = - i \partial_{\beta_\pm}$. Note that (\ref{qcon}) is multiplied by the factor $q^{-2}$ with respect to the Wheeler-DeWitt operator. Importantly, (\ref{qcon}) includes the extra term $\propto q^{-2}$. This repulsive potential is issued from a quantization consistent with the affine symmetry of the isotropic variables \cite{qb9a,qb9b}.   It is responsible for the avoidance of singularity in all the studied solutions. }

{In the present paper we focus on the operator (\ref{quham}). Note that it depends on the isotropic variable $q>0$ and so do its eigenstates. Therefore, the isotropic evolution can induce nonadiabatic transitions between anisotropy eigenstates.  This, however, is an issue of adiabatic and non-adiabatic approaches to quantum dynamics, which can be studied independently once the properties of  (\ref{quham}) are established. In what follows we derive  the asymptotic expressions for its spectrum.}

\subsection{The method}
The quantum numbers are denoted collectively by $\mathcal{I}$. Denoting the spectrum by $E^{(\mathcal{I})}_q$ we study  $\lim E^{(\mathcal{I})}_q$ as $q \to \infty$ and $q \to 0$. The method is based on the family of unitary dilations $U_{\xi(q)}$ {on $\mathcal{H}=L^2(\mathbb{R}^2,\ud\beta_+\ud\beta_-)$
$$ (U_{\xi(q)}\psi)\left(\bsb\right) := (\xi(q))^{-1} \psi\left(\frac{\bsb}{\xi(q)}\right)$$
 dependent on a function $\xi(q)$. When acting on $\widehat{\mathrm{C}}_{ani}(q)$ they leave the spectrum $E^{(\mathcal{I})}_q$ unchanged. } More precisely, we investigate the limits in $q$ of
$\widehat{\mathrm{C}}_{ani}^{(\xi)}(q)=U_{\xi(q)} \widehat{\mathrm{C}}_{ani}(q) U_{\xi(q)}^\dagger$
for some  $\xi(q)$  {that will be specified below}. The transformation $U_{\xi(q)}$ acts on $\hat{p}_{\pm}$ and $\hat{\beta}_\pm$ as
\begin{equation}
U_{\xi(q)} \hat{p}_{\pm} U_{\xi(q)}^\dagger = \xi(q)\, \hat{p}_{\pm}, \quad U_{\xi(q)} \hat{\beta}_\pm U_{\xi(q)}^\dagger = \frac{ 1}{ \xi(q)} \,\hat{\beta}_\pm\,.
\end{equation}
 {This leads to the unitarily equivalent Hamiltonian $\widehat{\mathrm{C}}_{ani}^{(\xi)}(q)$
\begin{equation}\label{rescon}
\widehat{\mathrm{C}}_{ani}^{(\xi)}(q) =  \dfrac{\xi(q)^2}{q^2}\,
\widehat{\mathrm{H}}(q)  \,,
\end{equation}
with
\begin{equation}
\label{dilatedH}
\widehat{\mathrm{H}}(q) =\hat{\mathbf{p}}^2+  \tilde{V}_q(\bsb),  \quad \tilde{V}_q(\bsb) = q^{8/3} \xi(q)^{-2} \, V(\bsb/\xi(q))\,.
\end{equation}
Choosing $\xi(q)$ as
\begin{equation}
\xi(q) = \dfrac{2}{3\,  \ln \left( 1+\dfrac{2}{3} q^{-2/3} \right)}\,,
\end{equation}
we prove in the sequel that the potential $ \tilde{V}_q(\bsb)$ in Eq. \eqref{dilatedH} possesses a well-defined limit for both small and large values of $q$, leading to an explicit spectrum of $\widehat{\mathrm{H}}(q=0)$ and $\widehat{\mathrm{H}}(q=\infty)$.  In other words the factor $q^{-2} \, \xi(q)^2$ in front of $\widehat{\mathrm{H}}(q)$ in the r.h.s of Eq. \eqref{rescon} captures both the divergent behavior (for small $q$) and {the vanishing behavior (for large $q$) of eigenvalues of $\widehat{\mathrm{C}}_{ani}^{(\xi)}(q)$.}

 {Note that we make use of $q$-dependent unitary transformations which couple to the isotropic evolution through the isotropic momentum operator in the constraint operator (\ref{qcon}). Although, the spectrum of the anisotropy operator is determined unambiguously, the obtained anisotropy eigenstates must be suitably rescaled before their use in a study of  (\ref{qcon}).}

\subsection{Harmonic behaviour for large values of $q$}
 {For large values of $q$ we have $\xi(q) \simeq q^{\frac{2}{3}}$. From the above we can see that the limit $q\to \infty$ corresponds to $\beta_{\pm}\to 0$ for the potential. The asymptotic expression for $\tilde{V}_q(\bsb)$ of Eq. \eqref{dilatedH} reads:
 \begin{equation}
\lim_{q \to \infty}  \tilde{V}_q(\bsb) = 8 (\beta_+^2+\beta_-^2) \,.
 \end{equation}
 Therefore
 \begin{equation}
 \widehat{\mathrm{H}}(+\infty) =  \hat{\mathbf{p}}^2 + 8 (\beta_+^2+\beta_-^2)\,.
 \end{equation}
Taking into account the scaling factor $q^{-2} \xi(q)^2$ of Eq. \eqref{dilatedH},  we conclude that the eigenvalues $E^{(\mathcal{I})}_q$ for large values of $q$ correspond to rescaled eigenenergies of a $2D$ isotropic harmonic oscillator and read
{\begin{equation}\label{harsp}
E^{(\mathcal{I})}_q  \underset{q \to \infty}{\simeq}   \frac{8}{q^{2/3}\sqrt{2}} \, \left(n_++n_- + 1 \right) + o(q^{-2/3}) \,,
\end{equation}
where the integers $n_\pm=0,1,\dots$ enumerate $1D$-harmonic oscillator energy levels.}
}

\subsection{Validity domain for the harmonic approximation}
\label{harmval}
 {Starting from the expression of $\widehat{\mathrm{C}}_{ani}(q)$ in Eq. \eqref{quham}, the equation with eigenvalue $E$ reads
\begin{equation}
\left( \frac{\hat{\mathbf{p}}^2}{q^{8/3}}+\, V(\bsb) - E \, q^{-2/3} \right) \psi_E(\bsb) = 0
\end{equation}
Since the eigenfunction $\psi_E$ is rapidly vanishing outside the domain $V(\bsb) - E \, q^{-2/3} \le 0$, the problem is well represented by a harmonic approximation, if in the domain $V(\bsb) - E \, q^{-2/3} \le 0$ the potential is essentially quadratic. This condition is valid for all $q$ because:\\
(a) For large $q$ the above condition reduces to the simple fact that $V(\bsb)$ is quadratic near the origin. As a matter of fact  we already know that the harmonic approximation holds true for large $q$. \\
(b) For small $q$ the kinetic energy term $\propto q^{-8/3}$ becomes dominant, and we know that kinetic energy is due to the oscillations of the wave function that takes place in the domain $V(\bsb) - E \, q^{-2/3} \le 0$. \\
A numerical analysis shows that $V(\bsb)$ is quadratic for $V(\bsb) \lesssim 1$. Therefore a harmonic approximation of the eigenvalues $E$ is validated if the following condition holds true
\begin{equation}
\label{harcond1}
E \, q^{-2/3} \lesssim 1 \,.
\end{equation}
This condition summarizes the intuitive breakdown of the harmonic approximation for large excitations and for small volumes. Using Eq. \eqref{harsp} the above condition can be translated into the following bound on the (harmonic) quantum { numbers $n_+$, $n_- $:
\begin{equation}
\label{harmcond2}
n_++n_-+1 \lesssim \dfrac{1}{4 \sqrt{2}} \, q^{4/3} \,.
\end{equation}}
This condition has consequences for modeling bouncing scenarios, as explained below in Sec. \ref{seccom}. { Let us stress that this condition holds irrespectively of adiabatic or nonadiabatic approximations applied to the full quantum dynamics and their validity.
}

\subsection{Steep wall behaviour for small values of $q$}
 {For small values of $q$ we have $\xi(q) \simeq |\ln q |^{-1}$ and we prove below that the asymptotic expression for $\tilde{V}_q(\bsb)$ in Eq. \eqref{dilatedH} reads:
\begin{equation}
\label{limpot}
\lim_{q \to 0}\,  \tilde{V}_q(\bsb)=V_\infty(\bsb), \,
\end{equation}
where $V_\infty$ is the infinite potential well corresponding to an equilateral triangular box with the side size $b=2/\sqrt{3}$. The potential $V_\infty$ is vanishing inside the triangle and infinite outside (except for three half-lines) as illustrated in Fig. \ref{figure2}.
}
\begin{figure}[!ht]
\includegraphics[width=0.4\textwidth]{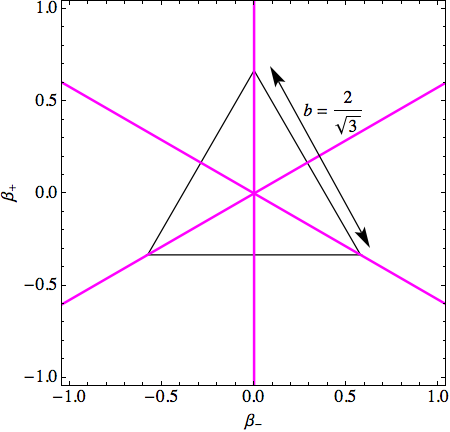}
\caption{The infinite potential well $V_\infty (\bsb)$ of  Eq.\;\eqref{limpot} corresponds to an equilateral triangular box. The three ${\sf C}_{3v}$ symmetry axes $\beta_-=0$, $\beta_+=\beta_-/\sqrt{3}$, $\beta_+=-\beta_-/\sqrt{3}$ are included. Inside the blue triangle $V_\infty(\bsb)$ is zero.}
\label{figure2}
\end{figure}

 {Because the potential $V(\bsb)$ possesses the $\mathsf{C}_{3v}$ symmetry (see Fig. \ref{figure1}),  it is sufficient to study the limit in Eq. \eqref{limpot} for $\beta_+ >0$. We first find  the equivalent
\begin{equation}
\forall \beta_+ >0, \, \forall \beta_- \ne 0, \, V( |\ln q| \bsb) \underset{q\to 0}{\simeq} \frac{1}{3} \, \exp \left[ 4 |\ln q| (\beta_++\sqrt{3} |\beta_-|) \right] \,.
\end{equation}
Therefore,
\begin{equation}
\nonumber
\forall \beta_+ > 0, \, \forall \beta_- \ne0\,, \,  \left\{ \begin{array}{cl}\text{if}\,\, \beta_++\sqrt{3} |\beta_-| > 2/3\,,& \lim_{q \to 0} q^{8/3} \ln^2 q\, V(|\ln q| \bsb) =+\infty \, , \\ \text{if}\,\, \beta_++\sqrt{3} |\beta_-| < 2/3\,, & \lim_{q \to 0} q^{8/3} \ln^2 q\, V(|\ln q| \bsb) =0 \,. \end{array}\right.
\end{equation}
We also find directly from the expression of $V$
\begin{equation}
\forall \beta_+ >0,\, \beta_-=0,\, \lim_{q \to 0} q^{8/3} \ln^2 q\, V(|\ln q| \bsb) =0 \,.
\end{equation}
Then, taking into account the $\mathsf{C}_{3v}$ symmetry of the potential, we construct the complete potential $V_\infty(\bsb)$ as represented in Fig. \ref{figure2}.
Having proved Eq. (\ref{limpot}) we rewrite the Hamiltonian $\widehat{\mathrm{H}}(q)$ of Eq. \eqref{dilatedH} for $q=0$ as
\begin{equation}
\widehat{\mathrm{H}}(0)=\hat{\mathbf{p}}^2+ V_\infty(\bsb)
\end{equation}
Up to a factor $1/2$ in front of $\hat{\mathbf{p}}^2$ in the above formula, the spectrum of this type of Hamiltonian is well-known \cite{WaiLi1985,WaiLi1987}\cite{Gaddah2013} and reads
\begin{equation}
e^{(T)}_{m,n} = \frac{8 \pi^2}{3 b^2} \left( \frac{m^2}{3} +
n^2+m n \right) =  \frac{8 \pi^2}{3 b^2} \left| n+ \frac{1}{\sqrt{3}} e^{i \pi/6} m \right|^2 \,,
\end{equation}
where $m=0,1,2,\dots$,  $n=1,2,\dots$, and $b=2/\sqrt{3}$. Taking into account the scaling factor $q^{-2} \xi(q)^2$ in Eq. \eqref{dilatedH}, we deduce that for small values of $q$ (and for fixed values of $m$ and $n$) the spectrum of $\widehat{\mathrm{C}}_{ani}(q)$ reads:
{
\begin{equation}
\label{steep}
E^{(\mathcal{I})}_q  \underset{q \to 0}{\simeq}  \frac{4 \pi^2 }{q^2 \ln^2 q} \left|  n  +  \frac{1}{\sqrt{3}} e^{i \pi/6} m\right|^2 + o(q^{-2} \ln^{-2} q)\,.
\end{equation}}
From Eq. (\ref{steep}) we deduce the limit
\begin{equation}
\label{dominant}
\lim_{q \to 0} q^2\, E^{(\mathcal{I})}_q = 0\,.
\end{equation}
The above property has significance for the singularity resolution, which we explain below.
}
\subsection{Comments}
\label{seccom}

 {First, our method shows in a straightforward way that it is possible to capture in a single factor $q^{-2} \xi(q)^2$ the principal part of the $q$-dependence of eigenvalues for large and small $q$. It  leads to a new Hamiltonian $\widehat{\mathrm{H}}(q)$ that possesses well-defined limits on both ends ($q=0$ and $q=\infty$). This opens the way toward future studies for a possible uniform approximation of eigenvalues.}

{Second, it is worth noting that the label $\mathcal{I}$ in Eq. \eqref{steep} is not an ordering parameter and the quantum numbers  $n$ and $m$} are different from those appearing  in the harmonic case ($n_\pm$) in  Eq. \eqref{harsp}. Therefore we cannot connect analytically both asymptotic expressions. Nevertheless, the ordering between eigenvalues for each limit ($q \to 0$ and $q \to \infty$) is meaningful and the $q$-dependence of the respective eigenenergies can be analysed.

 Third, the asymptotic Hamiltonians ($q \to 0$ and $q \to \infty$) do not possess a continuous spectrum. This constitutes a strong argument in support of the conjecture that $\widehat{\mathrm{C}}_{ani}(q)$ has no continuous spectrum for any value of $q$. The rigorous proof is given in Section \ref{proof}.  {The asymptotic analysis of eigenvalues alone does not give the threshold value $q_m$ that separates the two regimes of validity of the expressions given in Eqs \eqref{harsp} and \eqref{steep}. Yet a direct study as presented in Sec. \ref{harmval} gives the sought condition summarized by Eq. \eqref{harcond1}.
 }

Fourth, we have proved Eq. (\ref{dominant}). In the Misner paper \cite{cwm} where is introduced the quantum steep wall approximation, the coefficient $\ln^2 q$ of Eq. (\ref{steep}) is missing in the quantum energy formula and leads to the false idea that the quantum eigenenergies behave exactly as $\propto a^{-6}$ close to the singularity. This rough approximation has no qualitative consequence on the results of Misner's paper. However, in our previous papers  \cite{qb9a, qb9b} we have proved that the affine quantization of the isotropic dynamics given by the Hamiltonian constraint (\ref{condec}) produces a repulsive potential term $\propto q^{-2}$. Therefore, in our case this corrected dependence in $\ln^2 q$ is crucial as Eq. \eqref{dominant} implies that the repulsive potential is dominant close to the singularity.\footnote{ {In a completely different framework (supersymmetric model), with a different choice of coordinates, the same kind of bounce for a quantum Bianchi IX model can be found in \cite{ds}. }} It proves that a bounce must always exist in the Mixmaster model, independently of the harmonic approximation used in our previous papers. The harmonic approximation appears just as a simplified bouncing scenario (probably a smoother one), but the existence of a bounce itself is  unquestionable (at least in the adiabatic approximation). This point is crucial to validate our results in \cite{qb9a, qb9b} beyond the framework of the harmonic approximation.

 {Fifth, the inequality in Eq.\;\eqref{harmcond2} that specifies the domain of validity of the harmonic approximation has interesting consequences for bouncing models in general, and in particular for the one developed in our previous paper \cite{vib}. Indeed it proves the following: If the use of the harmonic approximation in a nonadiabatic framework leads to a dynamical behavior that does not violate \eqref{harmcond2} (at any time), then the harmonic approximation is sufficient to model the system (for the particular set of initial conditions that has been chosen).
In our case it validates the numerical simulations done in  \cite{vib} and then the conclusions of  that paper are also validated, namely the adiabatic behavior of low levels of excitations.}
\section{Discreteness of the spectrum}\label{proof}

\subsection{The criterion}
There exists in the mathematical literature a general  criterion for non-compact potentials to originate purely  discrete spectra. It was proved by Wang and Wu in 2008 \cite{wang-wu}. A clear account of  this result was later given by Simon in \cite{simon}.
These authors assert  that the Schr\"odinger operator in any dimension:
\begin{equation}
\hat{\mathrm{H}}=-\Delta+ V
\end{equation}
has a purely discrete spectrum if the Lebesgue  measure $| \cdot |$ of the projection set $\Omega_M(V)=\{x\, |\, 0\leq V(x)<M\}$ is finite: \begin{equation}
|\Omega_M(V)|<\infty.
\end{equation}
In the next section, we apply this criterion to prove that the spectrum of the Hamiltonian \eqref{quham} is purely discrete.

\subsection{Finiteness of the surface area}

Let us show that the surface area containing  points $\bsb=(\beta_+,\beta_-)$ satisfying
\begin{equation}
\Omega_M=\{\bsb\,: \,0\leq V(\bsb)<M\}
\end{equation}
is finite $|\Omega_M|<\infty$. In practice it needs to be shown that the area enclosed by the constant potential lines $V(\bsb)=M$ is finite. Several equipotential lines of (\ref{b9pot}) are plotted in Fig. (\ref{figure3}). They are closed for $M<1$ and open for $M\geqslant 1$. Thus, in order to prove the finiteness of $|\Omega_M|$ it is sufficient to consider the $M\geqslant 1$ case.
\begin{figure}[t]
\begin{center}
\includegraphics[width=.4\textwidth]{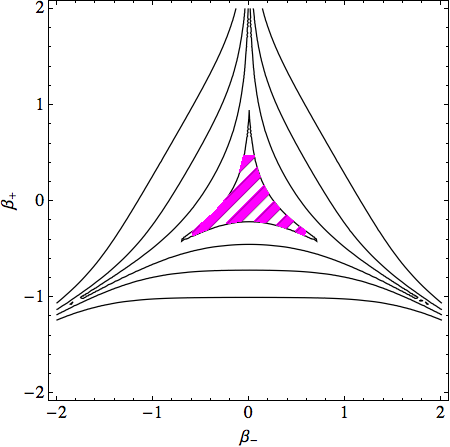}
\end{center}
\caption{Plot of the contours of the anisotropy potential $V(\bsb) = 0.8,~10,~10^2,~10^3$. The shaded region corresponds to the compact domain of $V(\bsb)<1$. The domain of $V(\bsb)<M$ is non-compact for $M\geqslant 1$ .}
\label{figure3}
\end{figure}

The enclosing curves satisfying $V(\bsb)=M\geqslant 1$ might  be parametrised  by the four following equations:
\begin{equation} \label{sp1}
\begin{split}
&\beta_-=\pm\frac{\sqrt{3}}{6}\arcosh\frac12\left(e^{-6\beta_+}+\sqrt{4+3(M-1)e^{-4\beta_+}} \right),~~\beta_+\in\mathbb{R}\\
&\beta_-= \pm\frac{\sqrt{3}}{6}\arcosh\frac12\left(e^{-6\beta_+}-\sqrt{4+3(M-1)e^{-4\beta_+}} \right),~~\beta_+\leqslant X\, ,
\end{split}
\end{equation}
where $X$ is the negative root of $ e^{-6\beta_+}-\sqrt{4+3(M-1)e^{-4\beta_+}}= 2$.
Due to the $\mathsf{C}_{3v}$ symmetry of the potential,  in order to prove that the enclosed surface area is finite,  it is sufficient to prove that the area of a part of the surface delimited by the curves \eqref{sp1}, say,
\begin{equation}\label{s1}
|\Omega_M(\beta_0)|=\frac{\sqrt{3}}{6}\int_{\beta_{0}}^{\infty}\arcosh\frac12\left(e^{-6\beta_+}+\sqrt{4+3(M-1)e^{-4\beta_+}} \right) \ud\beta_+
\end{equation}
is finite for some $\beta_0<\infty$. The surface $\Omega_M(\beta_0)$ for $M=100$ and $\beta_0=0$ is depicted in Fig. (\ref{figure4}).
\begin{figure}[t]
\begin{center}
\includegraphics[width=.4\textwidth]{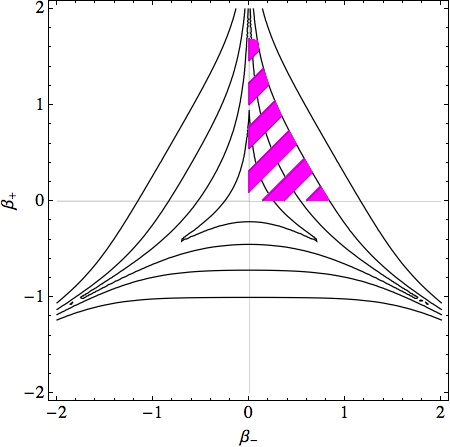}
\end{center}
\caption{The area of the shaded non-compact region $\Omega_{10^2}(0)$, enclosed by $\beta_-=0$, $\beta_+=0$, $\beta_-=\frac{\sqrt{3}}{6}\arcosh\frac12\left(e^{-6\beta_+}+\sqrt{4+297e^{-4\beta_+}} \right)$, is proved to be finite.}
\label{figure4}
\end{figure}
Let us estimate the area $|\Omega_M(\beta_0)|$ of Eq. (\ref{s1}) in a few steps. By making use of $\sqrt{1+x}\leq  1+\frac{x}2$ we get
\begin{align}\label{s2}
|\Omega_M(\beta_0)|<\frac{\sqrt{3}}{6}\int_{\beta_{0}}^{\infty}\arcosh\left(1+\frac12e^{-6\beta_+}+\frac{3(M-1)}{8}e^{-4\beta_+}\right) \ud\beta_+,
\end{align}
which for any $\beta_0>\ln\frac{\sqrt{3(M-1)}}{2}$ is further bounded by
\begin{align}\label{s3}
|\Omega_M(\beta_0)|<\frac{\sqrt{3}}{6}\int_{\beta_{0}}^{\infty}\arcosh\left(1+\frac{3(M-1)}{4}e^{-4\beta_+}\right) \ud\beta_+\, .
\end{align}
The application of the identity $\arcosh (x)\equiv \ln(x+\sqrt{x^2-1})$ and then twice the inequality $\sqrt{1+x}\leq  1+\frac{x}2$ gives:
\begin{equation}\label{s4}
\begin{split}
&|\Omega_M(\beta_0)|<\\ &\frac{\sqrt{3}}{6}\int_{\beta_{0}}^{\infty}\ln(1+\sqrt{\frac{3(M-1)}{2}}e^{-2\beta_+}+\frac{3(M-1)}{4}e^{-4\beta_+}+\left[\frac{3(M-1)}{8}\right]^{\frac32}e^{-6\beta_+}) \ud\beta_+\,.
\end{split}
\end{equation}
Since $\ln(1+x)\leq x$ we finally get
\begin{equation}\label{s5}
\begin{split}
&|\Omega_M(\beta_0)| <\\ &\frac{\sqrt{3}}{12}\left(\sqrt{\frac{3(M-1)}{2}}e^{-2\beta_0}+\frac{3(M-1)}{8}e^{-4\beta_0}+\frac13\left[\frac{3(M-1)}{8}\right]^{\frac32}e^{-6\beta_0}\right)<\infty\,.
\end{split}
\end{equation}
which completes the proof.

\section{Conclusion}
 {We have presented  several mathematical properties of the spectrum of the  Schr\"{o}dinger operator describing the anisotropic evolution of the Mixmaster model.
Our main result concerns the asymptotic expressions for the eigenenergies at large and small values of $q$. There are also established several interesting facts:}

 {First, a unique unitary transform is able to capture in a single factor the main $q$-dependence of eigenenergies.}

Second, the harmonic approximation used in  \cite{qb9a, qb9b} corresponds in fact to the mathematical asymptotic expression \eqref{harsp} for large values of $q$.

Third, the exact asymptotic behavior \eqref{steep} for small $q$ is not  the one given by Misner in \cite{cwm}: the factor $\ln^2 q$ is missing in Misner's formula. Then, thanks to the $\ln^2 q$ factor, Eq. \eqref{dominant} proves that the repulsive potential term  $\propto q^{-2}$ present in  \cite{qb9a, qb9b} is always dominant close to the singularity, even if the harmonic approximation is not valid. This point is crucial in validating our previous results on bouncing scenarios beyond the framework of the harmonic approximation.

 {Fourth, our asymptotic analysis of the spectrum for large $q$ complemented by a direct reasoning on the eigenfunctions is able to specify the limit on $q$ and $E$ that separates the two asymptotic regimes. }

 {Fifth, we have proved the discreteness of the spectrum despite the non-compact anisotropy potential. This result validates implementation of approximations of the potential, which remove the three non-compact canyons and lead to more manageable Schr\"{o}dinger operators.}

 {Finally, our analysis based on a unique unitary transform for all values of $q$ opens interesting perspectives in the search for the uniform approximation of eigenvalues.}

\section{Acknowledgments}
The authors are grateful  to Alain Joye (Univ. J. Fourier, Grenoble) for  pointing out the paper \cite{simon} {and anonymous referees for helping in improving the manuscript}.

\end{document}